\documentclass[%
 reprint,
 amsmath,amssymb,
 aps,
]{revtex4-1}

\usepackage{graphicx}
\usepackage{dcolumn}
\usepackage{bm}
\usepackage{color}

\newcommand{\ka}{{k_{\rm A}}}
\newcommand{\kb}{{k_{\rm B}}}

\begin{document}

\preprint{APS/123-QED}

\title{Cavity-based robustness analysis of interdependent networks: \\ Influences of intranetwork and internetwork degree--degree correlations }

\author{Shunsuke Watanabe}
 \email{watanabe@sp.dis.titech.ac.jp}
\author{Yoshiyuki Kabashima}%
 \email{kaba@dis.titech.ac.jp}
 \affiliation{
 Department of Computational Intelligence and Systems Science, Tokyo Institute of Technology, Yokohama 2268502, Japan
}

\
\date{\today}

\begin{abstract}
We develop a methodology for analyzing the percolation phenomena of
two mutually coupled (interdependent) networks based on the cavity
method of statistical mechanics. In particular, we take into account
the influence of degree--degree correlations inside and between the
networks on the network robustness against targeted (random degree-dependent) attacks and
random failures. We show that the developed methodology is reduced
to the well-known generating function formalism in the absence of
degree--degree correlations. The validity of the developed
methodology is confirmed by a comparison with the results of
numerical experiments. Our analytical results 
indicate that the
robustness of the interdependent networks depends 
on both the intranetwork and internetwork degree--degree correlations 
in a nontrivial way for both cases of random failures and targeted attacks. 

\end{abstract}

\pacs{Valid PACS appear here}
\maketitle


\section{\label{sec:level1}Introduction}
There are many kinds of complex systems 
in both natural and artificial worlds, and recently
these systems have come to be studied in various fields by being
handled as {\em networks}. Networks are expressed mathematically as
graphs in which the constituent elements and interactions among
these elements are expressed as sites (nodes, vertices) and bonds
(links, edges), respectively. Random networks in particular
\cite{Erdos, Bollobas}, which are randomly generated networks
characterized only by their macroscopic properties, have been widely
examined because of their analytical tractability.

One major concern surrounding networks is the robustness against
random failures (RFs) or targeted attacks (TAs). The size of the largest subset
of sites that are connected to one another, which is often referred
to as the giant component (GC) \cite{Jason}, generally becomes
smaller as more sites and/or bonds are removed. A standard measure
to characterize the network robustness is the critical rate of
failure at which the fraction of the GC against the size $N$ of the
original network vanishes from $O(1)$ to zero; this is often
referred to as the {\em percolation threshold} \cite{Kirkpatrick,
Callaway}. In general, a network becomes more tolerant to RFs/TAs as
each site in the network 
increases in {\em degree}, a parameter that
represents the number of bonds directly connected to the site. On
the other hand, increasing the number of bonds is generally costly
in terms of various aspects. Therefore, several earlier studies
examined the robustness of random networks that are specified by
only the degree distribution while keeping the average degree fixed
\cite{Valente,Paul}. However, the properties of real-world networks
cannot be fully characterized by the degree distribution. As a
logical step to take other statistical properties into account, the
influences of degree correlations between two directly connected
sites (degree--degree correlations) were also recently examined
\cite{Shiraki, Agliari, Tanizawa}.

More recently, a new type of model that considers {\em
interdependent networks} was proposed \cite{Buldyrev 2010, Parshani
PRL} and has attracted significant attention \cite{Gao, Son 2011, Morris, Baxter, Buldyrev 2011,  Liu, Zhou 2012, Schneider, Parshani EPL, Wang, Min, Cellai}. In this model, the system is composed of two
sub-networks in which the sites in one network are coupled with
those in the other on a one-to-one basis. The two sites of a pair
are dependent on each other, so that neither of the sites is
functional (active) when either is broken (inactive). This interdependence
between the two networks can facilitate a chain of failures, which
is sometimes referred to as {\em cascade phenomena}; removal of
sites in one network leads to the emergence of new isolated sites in
the other, which then acts as a trigger of new failures in the first
network and the process repeats itself. This mechanism can result in
guidelines for constructing a robust system that are different from
those known for single networks. For example, it is known that a
broad degree distribution makes a network more robust against random
site failures in the case of single networks \cite{Buldyrev 2011},
but in interdependent networks, the degrees in a network should be
uniform to increase the network robustness when the interdependent site pairs are randomly coupled between  the two networks because sites of a lower
degree tend to cause catastrophic breakdowns that amplify the
cascade phenomena \cite{Buldyrev 2010}. In a similar manner to the
single-network case, the influences of degree--degree correlations
in each network (intranetwork degree--degree correlations) have also
been examined numerically \cite{Zhou 2012}. However, as far as the
authors know, analytical examinations of the intracorrelations and
correlations between networks (internetwork degree--degree correlations)
have not yet been reported.

It is against this background that we develop here an analytical
methodology for investigating the influences of the intra- and
internetwork degree--degree correlations on the robustness of
interdependent networks. Our method is based on the statistical
mechanics cavity method developed for disordered systems
\cite{Mezard 1987,Mezard 2001, Mezard 2009} and evaluates the
probability that a pair of interdependent sites characterized by
their degrees belongs to the GC by utilizing the tree approximation
under the assumption that a GC of size $O(N)$ is formed.

The resultant methodology can be regarded as a generalization of the
well-known generating function formalism (GFF) \cite{Newman 2001} that systematically
evaluates various topological quantities by converting a graph into
a transcendental equation of a single variable. One can analytically
show that our methodology is reduced to the GFF in the absence of
any degree correlations; solving a set of nonlinear equations with
respect to multiple variables, which cannot be achieved with the
standard GFF, is indispensable for evaluating the size of the GC in
the presence of degree--degree correlations.

The remainder of this paper is organized as follows. In Sec.
\ref{preliminaries}, we briefly summarize the elemental concepts
necessary for our investigation. In Sec. \ref{section:
cavity_single}, we develop our analytical methodology on the basis
of the cavity method and discuss its relationship with the GFF. In
Sec. \ref{section: cavity_interdependent}, which is the main part
of this paper, we demonstrate how the developed method is applied to
interdependent networks, with the results for simple examples shown
in Sec. \ref{section: numerical_test}. We end the paper with a
summary.

\section{Preliminaries on networks}
\label{preliminaries}

We introduce here the definitions of several concepts that are
necessary for our analysis of interdependent networks. We denote the
degree distribution by $p(k)$, which indicates the fraction of sites
of degree $k$ in a network. Based on this, we can define the degree
distribution of a bond $r(k)$, which represents the probability that
one terminal of a randomly chosen bond has degree $k$:
\begin{equation}
r(k)=\frac{k p(k)}{\sum_l l p(l)}=\frac{kp(k)}{\langle k \rangle},
\end{equation}
where $\langle k \rangle$ is the average degree when a site in the
network is chosen randomly.

Although the degree distribution is a fundamental feature, it is not
sufficient to fully characterize the network properties \cite{Newman
2002}.
For instance, social networks exhibit the assortative tendency that
high-degree sites attach to other high-degree sites. In contrast,
technological and biological networks exhibit the disassortative
tendency that high-degree sites preferentially connect with
low-degree sites, and vice versa. To introduce such tendencies in a
simple manner, we characterize our network ensembles by utilizing
the joint degree--degree distribution $r(k,l)$, which is the
probability that the two terminal sites of a randomly chosen bond
have degrees $k$ and $l$. From this definition, one can relate
$r(k,l)$ to $p(k)$ and $r(k)$ as
\begin{equation}
\sum_l r(k,l) = r(k) =\frac{kp(k)}{\sum_l lp(l)},
\end{equation}
for $\forall k$.

Furthermore, the joint distribution is used to evaluate the
conditional distribution $r(k|l)$, i.e., the probability that one
terminal site of a randomly chosen bond has degree $k$ given that
the other terminal site has degree $l$, as
\begin{equation}
r(k|l)=\frac{r(k,l)}{r(l)}=\frac{\langle k \rangle r(k,l)}{l p(l)}.
\end{equation}
When the condition
\begin{equation}
r(k|l)=r(k)  \label{eq:r(k|l)=r(k)}
\end{equation}
holds for $\forall k,l$, the degrees of the directly connected sites
are statistically {\em independent}. To macroscopically quantify the
degree--degree correlations, the Pearson coefficient
\begin{equation}
C=\frac{1}{\sigma^2_r}\sum_{kl}kl(r(k,l)-r(k)r(l)),
\end{equation}
where
\begin{equation}
\sigma_r=\sum_{k}k^2r(k)-\left(\sum_{k}k r(k)\right)^2
\end{equation}
is often used. If $C$ is zero, then the random network is regarded
as {\em uncorrelated}. A positive (negative) $C$ indicates
assortative (disassortative) mixing.

In a pair of interdependent networks, labeled A and B, we assume
that each site in A is coupled with a site in B on a one-to-one
basis. We represent the probability that a randomly chosen
interdependent pair is composed of a site of degree $k_{\rm A}$ in A
and a site of degree $k_{\rm B}$ in B as $P(k_{\rm A},k_{\rm B})$.
Let us denote the joint degree--degree distribution for networks A
and B as $r_{\rm A}(\ka,l_{\rm A})$ and $r_{\rm B}(\kb,l_{\rm B})$,
respectively, For consistency, the identities
\begin{eqnarray}
p_{\rm A}(\ka)&=&\sum_{\kb} P(k_{\rm A},k_{\rm B}) \\
r_{\rm A}(\ka)&=&\sum_{l_{\rm A}}r_{\rm A}(\ka,l_{\rm
A})=\frac{\sum_{\kb} \ka P(\ka,\kb)}{ \sum_{\ka,\kb} \ka P(\ka,\kb)}
\label{rA}
\end{eqnarray}
for network A, and those for network B, must hold. Using Bayes'
formula, the conditional distribution that a site in A has degree
$\ka$ under the condition that the coupled site in B has degree
$\kb$ is evaluated as
\begin{eqnarray}
P_{\rm A}(\ka|\kb)=\frac{P(\ka,\kb)}{\sum_{\kb} P(\ka,\kb)},
\label{PA}
\end{eqnarray}
and
\begin{eqnarray}
\label{PB}
P_{\rm B}(\kb|\ka)=\frac{P(\ka,\kb)}{\sum_{\ka} P(\ka,\kb)}.
\end{eqnarray}
Equations (\ref{rA}) and (\ref{PB}) are 
used for assessing
the conditional
distributions for {\em site pairs}; namely, the probability that a
site pair of degrees $l_{\rm A}$ in A and $l_{\rm B}$ in B is
connected with another site pair of degrees $k_{\rm A}$ in A and
$k_{\rm B}$ in B by a link in A, which is evaluated as
\begin{eqnarray}
r_{\rm A}(k_{\rm A},k_{\rm B}|l_{\rm A},l_{\rm B})=P_{\rm B}(k_{\rm
B}|k_{\rm A})r_{\rm A}(k_{\rm A}|l_{\rm A}),
\label{pair_conditionalA}
\end{eqnarray}
and $r_{\rm B}(k_{\rm A},k_{\rm B}| l_{\rm A},l_{\rm B})= P_{\rm
A}(k_{\rm A}|k_{\rm B})r_{\rm B}(k_{\rm B}|l_{\rm B})$. These
conditional distributions play an important role in analyzing
interdependent networks.

In addition to this statistical characterization, we also handle a
single realization of randomly generated networks. To specify such a
network, we introduce the notation $\partial i$ for the set of all
adjacent sites that are connected directly to site $i$ and
$|\partial i|$ for the number of elements in $\partial i$. We use $X \backslash x$ to represent a set defined by
removing an element $x$ from the set $X$. Therefore, $\partial
i\backslash j$ refers to a set of sites that is defined by removing
site $j \in \partial i$ from $\partial i$. For a pair of
interdependent networks A and B, $\partial_{\rm A} i$ is used to
represent the set of site pairs that are linked directly to the site
pair $i$ in network A, with $\partial_{\rm B} i$ being that for
network B.

We can also represent a network as a bipartite graph. For this, we
denote site $i$ as a circle, an undirected link $a$ between two
sites $i$ and $j$ as a square, and make a link for a related circle
and square pair. This generally yields a bipartite graph in which
each square has two links, while the number of links connected to a
circle varies following a certain degree distribution
(Fig.~\ref{single_bipartite}). In the bipartite graph expression, we
denote $\partial a$ as the set of two circles connected to square
$a$ and $\partial i$ as the set of squares connected to circle $i$.
For a pair of interdependent networks A and B, $\partial_{\rm A} a$
and $\partial_{\rm A} i$ are used to represent the bipartite
graph expression of the connectivity of site pairs in network A, and
$\partial_{\rm B} a$ and $\partial_{\rm B} i$ are used for network
B.

\section{Cavity approach to single networks}
\label{section: cavity_single}

We review here the cavity approach to the robustness analysis of
single complex networks, which was developed in an earlier study
\cite{Shiraki}. The relationship with the GFF \cite{Newman 2001},
another representative technique in research on complex networks, is
also discussed.

\subsection{Message passing algorithm: microscopic  description}
\label{message passing: micro}

Let us suppose that a network that is sampled from an ensemble
characterized by $p(k)$ and $r(k,l)$ suffers from 
RFs or TAs. We employ the binary variable $s_i \in \{0,1\}$ to
represent whether site $i$ is active ($s_i=1$) or inactive ($s_i=0$)
due to the failure. To take the connectivity into account, we
introduce the state variable $\sigma_i \in \{0, 1\}$, which
indicates whether $\exists{j} \in \partial i$ belongs to the GC
$(\sigma_i=0)$ or does not $(\sigma_i=1)$
when $i$ is left out. 
Using these definitions,
$i$ is regarded as belonging to the GC if and only if
$s_i(1-\sigma_i)$ becomes unity, which gives the size of the GC as
\begin{eqnarray}
S=\frac{1}{N} \sum_{i=1}^N s_i (1-\sigma_i).
\label{sizeofGC}
\end{eqnarray}

Our analysis is based on the random-network property that the
lengths of the closed paths between two randomly chosen sites
(cycles) typically increase as $O(\ln N)$ as the size of the network
$N$ tends to infinity, as long as the variance of the degree
distribution is finite. This presumably holds even when the
degree--degree correlations are introduced and indicates that we can
locally handle a sufficiently large random network as a {\em tree}.

To incorporate this property in our analysis, we introduce the
concept of an {\em $i$-cavity system}, which is defined by removing
site $i$ from the original system. Let us define $m_{j \to i}=0$
when $\exists{h} \in \partial j$ belongs to the GC in the $i$-cavity
system, and $m_{j \to i}=1$, otherwise. Then, $\sigma_i$ vanishes if
and only if there exists a ${j} \in \partial i$ that is active
$(s_j=1)$ and has $m_{j\to i}=0$. This offers the basic equation
\begin{eqnarray}
\sigma_i=\prod_{j \in  \partial i} \left (1-s_j + s_j m_{j \to i} \right ).
\label{full_state}
\end{eqnarray}

Given an $i$-cavity system, we remove a site $j \in \partial i$ and
switch $i$ back on instead, which yields a $j$-cavity system. Then,
$m_{i\to j}$ vanishes if and only if $\exists{h} \in \partial i
\backslash j$ belongs to the GC in the $j$-cavity system. A general
and distinctive feature of trees is that when $i$ is removed,
$\forall{j} \in \partial i$ are completely disconnected with one
another. This indicates that $m_{i \to j}$ becomes unity if and only
if none of ${h} \in \partial i \backslash j$ belongs to the GC in the
$i$-cavity system. These definitions then provide the {\em cavity
equation}:
\begin{eqnarray}
m_{i \to j} =\prod_{h \in \partial i \backslash j} \left (1-s_h + s_h m_{h \to i} \right ).
\label{cavity_state}
\end{eqnarray}
This equation defined for all links over the network determines the
cavity variables $m_{j \to i}$ necessary for evaluating
Eq.~(\ref{full_state}) for every site $i$. 
Solving Eq.~(\ref{cavity_state}) by the method of iterative substitution 
given the initial condition of $m_{j\to i}=0$
and substituting
the obtained solution into Eq.~(\ref{full_state}) give the size of
the GC, Eq.~(\ref{sizeofGC}).

\subsection{Bipartite graph expression}
\label{subsection: graphical-expression}

\begin{figure}[htbp]
\begin{center}
\includegraphics[width=8cm]{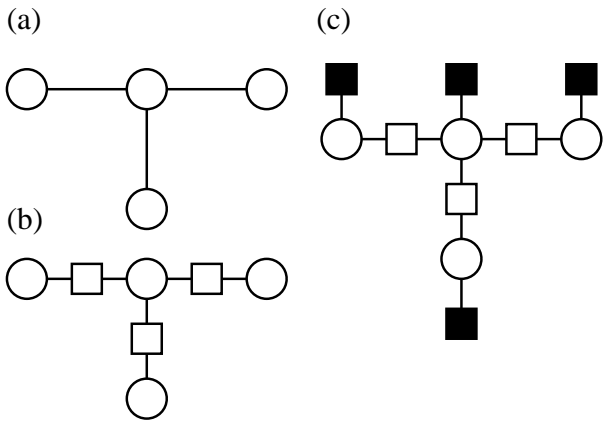}
\caption
{\label{single_bipartite} (a) Graph expression of a network. (b)
Bipartite graph expression of the graph in (a). A square node is
assigned for each edge in (a). (c) Bipartite graph expression of a
damaged network. Black square nodes attached to each circle
represent whether site $i$, denoted by the circle, is active
$(s_i=1)$ or inactive $(s_i=0)$.}
%
\end{center}
\end{figure}

In general, the cavity equations are expressed as message passing
algorithms on the bipartite graph corresponding to a given network
\cite{Mezard 1987}. For this, we denote $1-s_j+s_j m_{j \to i}$ in
two ways: $M_{j \to a}$ and $M_{a \to i}$, where $a$ represents a
square connected to two circles $i$ and $j$. Using these,
Eqs.~(\ref{cavity_state}) and (\ref{full_state}) can also be
expressed as
\begin{eqnarray}
M_{a \to i} &=&  M_{j \to a} \quad (\partial a =\{i,j\}) \label{h_step}\\
M_{i\to a} &=& 1-s_i +s_i \prod_{b \in \partial i \backslash a}M_{
b\to i },  \label{v_step}
\end{eqnarray}
and
\begin{eqnarray}
\sigma_i =\prod_{a \in \partial i}M_{a \to i},
\label{full_state2}
\end{eqnarray}
respectively.

One advantage of this expression is that the influence of $s_i$ can
be expressed graphically as an additional square node attached to
circle $i$ (Fig.~\ref{single_bipartite} (c)), which enables us to
interpret Eqs.~(\ref{h_step}) and (\ref{v_step}) as an algorithm
that computes an outgoing message along a link from a node on the
basis of incoming messages along the other links to the node. This
type of interpretation is useful for constructing cavity equations
to handle more advanced settings such as interdependent networks.

\subsection{Macroscopic description}
\label{subsection: single-macro}

We turn now to an evaluation of the typical size of the GC when the
networks are generated from an ensemble and damaged by RFs or TAs.
We assume that, as a consequence of the failures, each site of
degree $l$ is active only with a degree dependent probability $q_l$.
We employ the bipartite graph expression, classify every site by its
degree $l$, and define $I_l$ to be the frequency that sites of
degree $l$ receive $M_{a\to i}=1$ among all links of the bipartite
graph. Namely, $I_l$ is evaluated as $I_l=\left (l \sum_{i
}\delta_{|\partial i|,l}\right )^{-1} \sum_i (\delta_{|\partial
i|,l}\sum_{a \in \partial i} M_{a\to i})$, where the denominator is
the number of links that connect to sites of degree $l$, and the
numerator is the number of links for which $M_{a\to i}=1$ is
received by sites of degree $l$.

Although $I_l$ has sample-to-sample fluctuations that depend on the
network realizations, the strength of the fluctuations tends to zero
as $N$ becomes larger. For typical samples, this indicates that the
samples are expected to converge to their average as $N \to \infty$,
a property known as \textit {self-averaging} \cite{Mezard 1987}. In
the current problem, because of the tree-like property of the random
networks, $I_l$ can be evaluated as the expectations with respect to
the graph and failure generations. In the bipartite graph expression,
given a circle of degree $l$, the conditional distribution that 
the circle is coupled to a circle of degree $k$ through a square is given by
$r(k|l)$. In addition, the probability that the circle of degree $k$
is active is $q_k$. Averaging Eq.~(\ref{v_step}) with respect to
$r(k|l)$ and $q_k$ for a fixed value of $l$ and utilizing
Eq.~(\ref{h_step}) leads to
\begin{eqnarray}
I_l=\sum_{k} r(k|l) (1-q_k + q_k I_k^{k-1}).
\label{h_step2}
\end{eqnarray}
Here, we have employed the property that the average of $\prod_{b
\in \partial i \backslash a}M_{ b\to i }$ on the right-hand side of
Eq.~(\ref{v_step}) can be taken independently of the indices $b \in
\partial i \backslash a$ because of the tree-like nature of random
graphs.
The whole set in Eq.~(\ref{h_step2}) determines $I_l$ for
$\forall{l}$. After solving the equations, the typical size of the
GC is evaluated as
\begin{eqnarray}
\mu =\sum_{l} p(l) q_l (1-I_l^l),
\label{typical_GC}
\end{eqnarray}
which corresponds to Eq.~(\ref{sizeofGC}).

Equation (\ref{h_step2}) always allows a trivial solution $I_l=1$
for $\forall{l}$ yielding $\mu=0$. The local stability of this
trivial solution can be evaluated by linearizing the equations,
which gives
\begin{eqnarray}
\delta I_l= \sum_{k} (k-1)q_k r(k|l) \delta I_k,
\label{linear_stability}
\end{eqnarray}
or the alternative expression
\begin{eqnarray}
\delta {\bf I} ={\bf A} \delta {\bf I},
\label{vector_linear_stability}
\end{eqnarray}
where $\bf A$ is a matrix defined as
\begin{eqnarray}
A_{lk}=(k-1)q_k r(k|l).
\label{Alk}
\end{eqnarray}
Equation (\ref{vector_linear_stability}) states that the trivial
solution is stable provided $\mu=0$ if and only if all eigenvalues
of $\bf A$ are placed inside the unit circle centered at the origin
in the complex plane. 
As $A_{lk} \ge 0$ is guaranteed for $\forall{l}$ and $\forall{k}$, the Perron-Frobenius theorem indicates that 
the critical condition changing this situation
is given as
\begin{eqnarray}
{\rm det} \left ({\bf E} - {\bf A} \right )=0,
\label{critical_percolation}
\end{eqnarray}
which determines the percolation threshold for a given set of
control parameters, where $\bf E$ is the identity matrix.

When the active probabilities $q_l$ are sufficiently small for
$\forall{l}$, the absolute values of the eigenvalues of $\bf A$
become so small that the trivial solution is guaranteed to be
stable, yielding a vanishingly small GC size of $\mu=0$. However, as
the values of $q_l$ become larger in a certain manner,
Eq.~(\ref{critical_percolation}) is satisfied at the percolation
threshold, and a solution of $\mu>0$ comes continuously from the
trivial solution. In this way, the emergence of a large GC is always
described as a {\em continuous} phase transition for single
networks.

\subsection{Connection between the cavity method and generating function formalism}

Before proceeding further, we mention briefly the relationship
between the cavity method and the GFF.

Consider the cases of no degree correlations $r(k|l)=r(k)$ for
arbitrary pairs of ${k,l}$ and no site dependence of the active
probability $q_l=q$ for $\forall{l}$. In such cases,
Eq.~(\ref{h_step2}) becomes independent of $l$, and therefore we can
set $I_l=I$. This makes it possible to summarize Eq.~(\ref{h_step2})
as
\begin{eqnarray}
\sum_{l} r(l) I^{l-1} =\sum_{l}r(l) \left (1-q+q \sum_{k} r(k) I^{k-1} \right )^{l-1},
\label{GFF1}
\end{eqnarray}
which can be expressed more concisely as
\begin{eqnarray}
f=H(1-q+qf),
\label{GFF2}
\end{eqnarray}
where we have defined
\begin{eqnarray}
G(x)&=&\sum_k p(k) x^k,\label{eq:G(x)}\\
H(x)&=&\sum_k r(k) x^{k-1}=\frac{G'(x)}{G'(1)}\label{eq:H(x)},
\end{eqnarray}
and set $f\equiv \sum_{k} r(k)I^{k-1}=H(I)$. Using the solution of
Eq.~(\ref{GFF2}), Eq.~(\ref{typical_GC}) is evaluated as
\begin{eqnarray}
\mu=q \left (1 - G(1-q+qf) \right ).
\label{GFF3}
\end{eqnarray}

Equation (\ref{GFF2}) is nothing but a {\em transcendental equation}
for the GFF \cite{Buldyrev 2011}. Namely, the cavity method is
reduced to the GFF in the simplest cases. However, when
degree--degree correlations exist, $I_l$ generally depends on the
degree $l$, and therefore the cavity equations, Eq.~(\ref{h_step2}),
cannot be summarized as a nonlinear equation of a single variable.
As a consequence, one cannot exploit the compact expression of the
GFF and has to directly deal with the cavity equations for
evaluating the size of the GC \cite{Shiraki, Tanizawa}.
A similar idea has been implemented in the GFF by handling a set of
coupled generating functions for evaluating the GC of networks free
from failures \cite{Newman 2002}.


\section{Cavity approach for interdependent networks}
\label{section: cavity_interdependent} In this section, we develop
the cavity method for analyzing the cascade phenomena in
interdependent networks that occurs as a result of RFs or TAs.

\subsection{Cascade phenomena of interdependent networks}
Consider the pair of interdependent networks introduced in Sec.
\ref{preliminaries}. Each pair of sites in networks A and B is
interdependent so that both sites become inactive and lose their
functions if one site becomes inactive. In addition, each active
site in A also loses its function if it is disconnected from the GC
of A, which brings about functional failure of the coupled site in
B, and vice versa. In each network, the GC is defined as the largest
subset of functional sites.

We assume initial conditions of all sites being active and
functional in both networks. In the initial step, sites in A suffer
from RFs or TAs, and only a portion of the sites 
remain active.
Further, an additional portion of sites 
lose their functions
because they were disconnected from the GC of A. In the second step,
the sites in B that are coupled with the sites in A that were
disconnected from the GC also lose their functions due to the
properties of interdependent networks noted above. This reduces the
size of the GC in B, and an extra potion of sites in B lose their
functions. In the third step, functional failure in B is propagated
back to A causing further functional failure in A, and this process
is repeated until convergence. This is the cascade phenomena.

At convergence, every site of the GC in A is coupled with a site of
the GC in B on a one-to-one basis. The resulting GC of the site
pairs is often termed the {\em mutual GC}. Earlier studies reported
that unlike the case of single networks, the size of the mutual GC
relative to the network size vanishes discontinuously from $O(1)$ to
zero at a critical condition as the strength of the failures in the
initial step becomes larger \cite{Buldyrev 2010, Parshani
PRL}. We develop here a
methodology for examining this phenomena on the basis of the cavity
method by taking degree--degree correlations into account.

\subsection{Microscopic description}

\begin{figure}[htbp]
\begin{center}
\includegraphics[width=8cm]{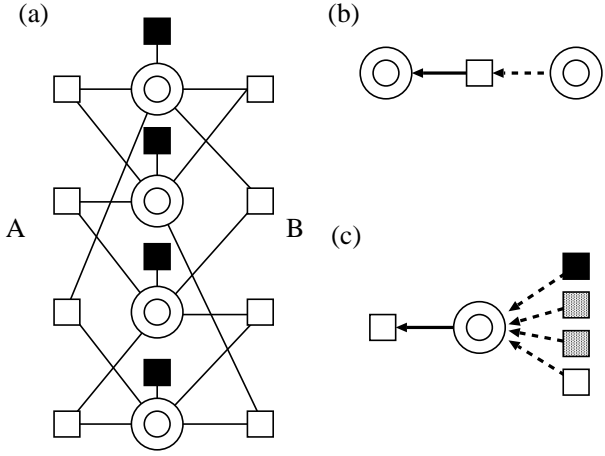}
\caption
{ \label{interdependent_bipartite} (a) Bipartite graph expression of
interdependent networks. Double circles represent site pairs
connected by internetwork links. White squares represent links in
each network, while each black square represents 
whether
the site in A of the site pairs is removed in the initial failure or not. 
(b) Graphical
expression of Eq.~(\ref{hstep_ID_A}) in A and Eq.~(\ref{hstep_ID_B})
in B. (c) Graphical expression of Eq.~(\ref{vstep_ID_A}) in A and
Eq.~(\ref{vstep_ID_B}) in B. Shaded squares indicate square nodes in
the counterpart network, messages from which are evaluated in the
previous step and fixed in the current update. Messages from the
black squares are fixed as $s_i$ in all steps. Contributions from
the shaded and black squares are summarized as $\tau_{{\rm
A},i}^{2t-1}$ and $\tau_{{\rm B},i}^{2t}$ in Eq.~(\ref{vstep_ID_A})
and Eq.~(\ref{vstep_ID_B}), respectively. In (b) and (c), the
message of the full line is computed from message(s) of the broken
line(s).
}
\end{center}
\end{figure}

Let us denote a site pair of the interdependent networks as $i$. We
employ a binary variable $s_i \in\{0,1\}$ to represent whether $i$
is kept active in the initial stage $(s_i=1)$ or fails $(s_i=0)$. We
also introduce the state variable $\sigma_{{\rm A},i}^t \in \{0,1\}$
that indicates whether $\exists{j} \in \partial_{\rm A} i$ belongs
to the GC in A $(\sigma_i^{\rm A}=0)$ or does not $(\sigma_i^{\rm
A}=1)$ 
when $i$ is left out
after the $t$-th $(t=1,2,\ldots)$ step, with $\sigma_{{\rm
B},i}$ being the equivalent state variable for B. Using these, the
size of the mutual GC after the $2t$-th step is expressed as
\begin{eqnarray}
S^{2t}=\frac{1}{N} \sum_{i=1}^N s_i(1-\sigma_{{\rm A},i}^{2t-1})(1-\sigma_{{\rm B},i}^{2t}).
\label{mutualGC_micro}
\end{eqnarray}

For a given $t$, $\sigma_{{\rm A},i}^{2t-1}$ and $\sigma_{{\rm
B},i}^{2t}$ can be obtained by the cavity method. To do this, we
note that the state variables in A after the $2t-1$-th step can be
evaluated by the scheme for single networks by handing $\tau_{{\rm
A},i}^{2t-1}=s_i (1-\sigma_{{\rm B},i}^{2t-2})$ as a binary variable
for representing whether $i$ is active $(\tau_{{\rm A},i}^{2t-1}=1)$
or inactive $(\tau_{{\rm A},i}^{2t-1}=0)$, where $\sigma_{{\rm
B},i}^{0}$ is set to zero from the assumption. Using the bipartite
graph expression (Fig.~\ref{interdependent_bipartite}), we get
\begin{eqnarray}
M_{a \to i}^{\rm A} &=& M_{j \to a}^{\rm A} \quad \left (\partial_{\rm A} a =\{i,j\} \right ), \label{hstep_ID_A} \\
M_{i \to a}^{\rm A} &=&  \!1\!- \! \tau_{{\rm A},i}^{2t-1} \! + \!
\tau_{{\rm A},i}^{2t-1} \! \prod_{b \in \partial_{\rm A} i
\backslash a} M_{b \to i }^{\rm A} , \label{vstep_ID_A}
\end{eqnarray}
and the solution of these yields
\begin{eqnarray}
\sigma_{{\rm A},i}^{2t-1} =\prod_{a \in \partial_{\rm A} i} M_{a \to i}^{\rm A}.
\label{full_ID_A}
\end{eqnarray}
Similarly, those in B after the $2t$-th step are
\begin{eqnarray}
M_{a \to i}^{\rm B} &=& M_{j \to a}^{\rm B} \quad \left (\partial_{\rm B} a =\{i,j\} \right ), \label{hstep_ID_B} \\
M_{i \to a}^{\rm B} &=&  \!1\!- \! \tau_{{\rm B},i}^{2t} \! + \! \tau_{{\rm B},i}^{2t} \!
\prod_{b \in \partial_{\rm B} i \backslash a} M_{b \to i }^{\rm B} ,
\label{vstep_ID_B}
\end{eqnarray}
and
\begin{eqnarray}
\sigma_{{\rm B},i}^{2t} =\prod_{a \in \partial_{\rm B} i} M_{a \to i}^{\rm B},
\label{full_ID_B}
\end{eqnarray}
where we set $\tau_{{\rm B},i}^{2t}=s_i (1-\sigma_{{\rm
A},i}^{2t-1})$. Solving these with the initial conditions $M_{i\to
a}^{\rm A}=0$ or $M_{i\to a}^{\rm B}=0$ at each step and inserting
the resultant values of Eqs.~(\ref{full_ID_A}) and (\ref{full_ID_B})
into Eq.~(\ref{mutualGC_micro}) offer the size of the GC after the
$2t$-th step for a given sample of interdependent networks and
initial failures.

\subsection{Macroscopic description}
To evaluate the typical size of the mutual GC when the
interdependent networks are generated as mentioned in Sec.
\ref{preliminaries} and suffer from RFs or TAs, we assume that each
site pair of degrees $l_{\rm A}$ in A and $l_{\rm B}$ in B is kept
active with a degree pair dependent probability $q_{l_{\rm A}l_{\rm
B}}$ as a consequence of the failures
that are brought about in network A at the initial stage. 
For RFs, we set
\begin{eqnarray}
\! q_{l_{\rm A}l_{\rm B}} \!=q, 
\end{eqnarray}
which implies that failures are generated randomly irrespectively of the values of $l_{\rm A}$ and $l_{\rm B}$. 
On the other hand, TAs are made preferentially for sites of the larger degrees in network A. 
Therefore, for a given value of the average active rate $q$, we assign the values of $q_{l_{\rm A}l_{\rm B}} $ as
\begin{eqnarray}
q_{l_{\rm A}l_{\rm B}} = \left \{
\begin{array}{ll}
0 & (l_{\rm A}> \Theta) \cr
\Delta & (l_{\rm A}= \Theta), \\
1 & (l_{\rm A}< \Theta) 
\end {array}
\right . 
\end{eqnarray}
where $\Theta$ and $\Delta$ are uniquely determined so that 
\begin{equation}
q = \sum_{l_{\rm B}} \left (\Delta P(\Theta, l_{\rm B})+\sum_{l_{\rm A} < \Theta } P(l_{\rm A}, l_{\rm B}) \right )
\end{equation}
holds. 

The key idea of our analysis is basically the same as that for single networks; namely, we
describe the system using conditional frequencies that a site pair
characterized by a pair of degrees $l_{\rm A}$ in A and $l_{\rm B}$
in B receives messages of unity from connected links in
A and B. For this, we define $I^{\rm A}_{l_{\rm A}l_{\rm B}} \!=\!
\left (l_{\rm A} \! \sum_{i} \delta_{|\partial_{\rm A} i|,l_{\rm
A}}\delta_{|\partial_{\rm B} i|,l_{\rm B}} \right )^{-1} \! \left
(\! \sum_{i} \! \delta_{|\partial_{\rm A} i|,l_{\rm A}}\! \sum_{a
\in \partial i}\! M_{a\to i}^{\rm A}\!  \right )$,
and similarly define $I^{\rm B}_{l_{\rm A}l_{\rm B}}$ for B.

Let us denote $q_{{\rm A},l_{\rm A}l_{\rm B}}^{2t-1}$ as the
conditional probability that $\tau_{{\rm A},i}^{2t-1}$ takes a value
of unity for site pairs of degrees $l_{\rm A}$ in A and $l_{\rm B}$
in B at the $2t-1$-th step. The self-averaging property and the
tree-like nature of random graphs allow us to macroscopically
describe Eqs.~(\ref{hstep_ID_A}) and (\ref{vstep_ID_A}) as
\begin{eqnarray}
&& I^{\rm A}_{l_{\rm A}l_{\rm B}}=\! \sum_{k_{\rm A},k_{\rm B}}  \!
r_{\rm A} \!( k_{\rm A}, \! k_{\rm B}|l_{\rm A}, \! l_{\rm B}) \cr
&& \hspace*{1.5cm} \times \left (\! 1\! - \! q_{{\rm A},k_{\rm
A}k_{\rm B}}^{2t\!-\!1} \! +\! q_{{\rm A},k_{\rm A}k_{\rm
B}}^{2t\!-\!1} (I^{\rm A}_{k_{\rm A} k_{\rm B}} )^{k_{\rm A}-1}\!
\right ). \label{macroh_stepA}
\end{eqnarray}
Equation (\ref{full_ID_A}) states that the conditional probability
of a site pair of degrees $l_{\rm A}$ in A and $l_{\rm B}$ in B
having $\sigma_{{\rm A},i}^{2t-1}=1$ after the $2t-1$-th step is
$\left (I^{\rm A}_{l_{\rm A}l_{\rm B}} \right )^{l_{\rm A}} $. Among
these sites, only the fraction of $q_{l_{\rm A}l_{\rm B}}$ is active.
Therefore, the conditional probability that a site pair of degrees
$l_{\rm A}$ in A and $l_{\rm B}$ in B has $\tau_{{\rm B},i}^{2t}=1$
in B at the $2t$-th step is evaluated as
\begin{eqnarray}
q_{{\rm B},l_{\rm A} l_{\rm B}}^{2t}=
q_{l_{\rm A}l_{\rm B}}\left (1-\left (I^{\rm A}_{l_{\rm A}l_{\rm B}} \right )^{l_{\rm A}} \right ),
\label{q2tB}
\end{eqnarray}
using the solution of Eq.~(\ref{macroh_stepA}). Equation
(\ref{q2tB}) makes it possible to macroscopically describe the
cavity equation in B at the $2t$-th step in a similar manner to
Eq.~(\ref{macroh_stepA})  as
\begin{eqnarray}
&& I^{\rm B}_{l_{\rm A}l_{\rm B}}=\! \sum_{k_{\rm A},k_{\rm B}}  \!
r_{\rm B} \!( k_{\rm A}, \! k_{\rm B}|l_{\rm A}, \! l_{\rm B}) \cr
&& \hspace*{1.5cm} \times \left (\! 1\! - \! q_{{\rm B},k_{\rm
A}k_{\rm B}}^{2t} \! +\! q_{{\rm B},k_{\rm A}k_{\rm B}}^{2t} (I^{\rm
B}_{k_{\rm A} k_{\rm B}})^{k_{\rm B}-1} \! \right ).
\label{macroh_stepB}
\end{eqnarray}
Using the solution of Eq.~(\ref{macroh_stepB}), the conditional
probability that a site pair of degrees $l_{\rm A}$ in A and $l_{\rm
B}$ in B has $\tau_{{\rm A},i}^{2t+1}=1$ in A at the $2t+1$-step is
evaluated as
\begin{eqnarray}
q_{{\rm A},l_{\rm A} l_{\rm B}}^{2t+1}=
q_{l_{\rm A}l_{\rm B}}\left (1-\left (I^{\rm B}_{l_{\rm A}l_{\rm B}} \right )^{l_{\rm B}} \right ).
\label{q2tplus1A}
\end{eqnarray}
Equation (\ref{mutualGC_micro}) gives the expectation of the size of
the mutual GC after the $2t$-th step as
\begin{eqnarray}
&& \mu^{2t}= \sum_{l_{\rm A},l_{\rm B}} P(l_{\rm A},l_{\rm B}) q_{l_{\rm A}l_{\rm B}}
\left (1-\left (I^{\rm A}_{l_{\rm A}l_{\rm B}} \right )^{l_{\rm A}} \right ) \cr
&& \hspace*{2cm} \times
\left (1-\left (I^{\rm B}_{l_{\rm A}l_{\rm B}} \right )^{l_{\rm B}} \right ) \cr
&& \phantom{\mu^{2t}}= \sum_{l_{\rm A},l_{\rm B}} P(l_{\rm A},l_{\rm B}) \left (q_{l_{\rm A}l_{\rm B}} \right )^{-1}
q_{{\rm A},l_{\rm A} l_{\rm B}}^{2t-1} q_{{\rm B},l_{\rm A} l_{\rm B}}^{2t}.
\label{mutualGC2t}
\end{eqnarray}
Equations (\ref{macroh_stepA})--(\ref{mutualGC2t}) constitute the
main result of the present paper.

One thing is noteworthy here. Similar to the case of single
networks, Eq.~(\ref{macroh_stepA}) always allows a trivial solution
of $I_{l_{\rm A}l_{\rm B}}^{\rm A}=1$, which becomes stable when
sufficiently small $q_{{\rm A}, l_{\rm A}l_{\rm B}}^{2t-1}$ values
are set for all pairs of $l_{\rm A}$ and $l_{\rm B}$. This yields
$q_{{\rm B}, l_{\rm A}l_{\rm B}}^{2t}=0$ for all pairs of $l_{\rm
A}$ and $l_{\rm B}$ in Eq.~(\ref{q2tB}), which makes $I_{l_{\rm
A}l_{\rm B}}^{\rm B}=1$ the unique and stable solution of
Eq.~(\ref{macroh_stepB}) at the $2t$-th step, offering $\mu^{2t}=0$.
In addition, at the subsequent $2t+1$-th step, $q_{{\rm A}, l_{\rm
A}l_{\rm B}}^{2t+1}=0$ holds for all pairs of $l_{\rm A}$ and
$l_{\rm B}$, which once more guarantees that the trivial solution is
stable. This means that unlike the case of single networks, the
trivial solution of $\mu^*=\lim_{t \to \infty} \mu^{2t} =0$ is
always locally stable in the dynamics of
Eqs.~(\ref{macroh_stepA})--(\ref{q2tplus1A}) irrespective of the
strength of the RFs. A finite $\mu^*$ is obtained when sufficiently
large $q_{{\rm A},l_{\rm A}l_{\rm B}}^{1}=q_{l_{\rm A} l_{\rm B}}$
values are set in the initial step for all pairs of $l_{\rm A}$ and
$l_{\rm B}$. As a consequence, the transition of a finite value of
$\mu^*$ to zero {\em generally} occurs in a {\em discontinuous}
manner for interdependent networks even when degree--degree
correlations are taken into account. Earlier studies, however, have
already reported the occurrence of the discontinuous transition for
a few specific examples \cite{Buldyrev 2010, Buldyrev 2011, Zhou
2012}.

\subsection{Relationship with earlier studies}
In Ref. \cite{Buldyrev 2011}, the case involving the highest internetwork
degree--degree correlation ($P(k_{\rm A},k_{\rm B})=\delta_{k_{\rm
B},k_{\rm A}}p(k_{\rm A})$) and no intranetwork degree--degree
correlation ($r_{\rm A}(k,l)=r_{\rm B}(k,l)=r(k)r(l)$) is examined
for degree-independent RFs characterized by $q_{l_{\rm A} l_{\rm
B}}=q$. In such a case, we can assume that $I_{l_{\rm A}l_{\rm
B}}^{\rm A}=I_{l_{\rm A}l_{\rm B}}^{\rm B}=I$, ignoring the
dependence on the degree and network. Further, we can set $\lim_{t
\to \infty} q_{{\rm A},ll}^{2t-1}= \lim_{t \to \infty} q_{{\rm
B},ll}^{2t}=q(1-I^l)$ because of the symmetry between networks A and
B. Inserting these into Eq.~(\ref{macroh_stepA}), in conjunction
with $r_{\rm A}(k_{\rm A}, k_{\rm B}|l_{\rm A},l_{\rm
B})=\delta_{k_{\rm B},k_{\rm A}}r(k_{\rm A})\delta_{l_{\rm A},l_{\rm
B}}$, we obtain an equation concerning $I$:
\begin{eqnarray}
I&=&1-q\left (1-IH(I) \right )+q \left (H(I)-IH(I^2) \right ) \cr
&=&1-q\left (1-(I+1)H(I)+IH(I^2) \right ),
\label{Cwilich}
\end{eqnarray}
which is equivalent to Eq.~(36) in Ref. \cite{Buldyrev 2011}. The size of
the mutual GC for $t\to \infty$ is evaluated by inserting $I_{l_{\rm
A}l_{\rm B}}^{\rm A}=I_{l_{\rm A}l_{\rm B}}^{\rm B}=I$ into
Eq.~(\ref{mutualGC2t}) for $P(k_{\rm A},k_{\rm B})=\delta_{k_{\rm
B},k_{\rm A}}p(k_{\rm A})$. This provides the expression
\begin{eqnarray}
\mu^*=q \left (1-2G(I)+G(I^2) \right ),
\label{Cwilich2}
\end{eqnarray}
which is also equivalent to Eq.~(35) in Ref. \cite{Buldyrev 2011}.

In Ref. \cite{Buldyrev 2010}, the case of no intranetwork degree--degree
correlation ($r_{\rm A}(k,l)=r_{\rm A}(k)r_{\rm A}(l)$, $r_{\rm
B}(k,l)=r_{\rm B}(k)r_{\rm B}(l)$) and no internetwork
degree--degree correlation ($P(k_{\rm A},k_{\rm B})=p_{\rm A}(k_{\rm
A})p_{\rm B}(k_{\rm B})$) is discussed. In such a case, one can set
$I_{l_{\rm A}l_{\rm B}}^{\rm A}=I_{\rm A}$ and $I_{l_{\rm A}l_{\rm
B}}^{\rm B}=I_{\rm B}$ by ignoring the site dependence. Let us focus
on the case of degree independent RFs $q_{l_{\rm A} l_{\rm B}}=q$
and the convergent state. Inserting $r_{\rm A}(k_{\rm A},k_{\rm
B}|l_{\rm A},l_{\rm B}) =p_{\rm B}(k_{\rm B}) r_{\rm A}(k_{\rm A})$
into Eq.~(\ref{macroh_stepA}) yields
\begin{eqnarray}
I_{\rm A}&=&1-q_{\rm B} +q_{\rm B} \left (\sum_{k_{\rm A}} r_{\rm
A}(k_{\rm A})I_{\rm A}^{k_{\rm A}-1}  \right )\cr &=&1-q_{\rm B}
+q_{\rm B} f_{\rm A}, \label{I2fnature}
\end{eqnarray}
or alternatively
\begin{eqnarray}
f_{\rm A}=H_{\rm A}(1-q_{\rm B}+q_{\rm B} f_{\rm A} ),
\label{nature1}
\end{eqnarray}
where $H_{\rm A}(x)=\sum_{k} r_{\rm A}(k)x^{k-1}$ and
$f_{\rm A}=H_{\rm A}(I_{\rm A})$.
In a similar manner, Eq.~(\ref{macroh_stepB}) offers
\begin{eqnarray}
f_{\rm B}=H_{\rm B}(1-q_{\rm A}+q_{\rm A} f_{\rm B}),
\label{nature2}
\end{eqnarray}
where $H_{\rm B}(x)=\sum_{k} r_{\rm B}(k)x^{k-1}$. Equations
(\ref{q2tplus1A}) and (\ref{q2tB}) provide $q_{\rm A}$ and $q_{\rm
B}$ in Eqs.~(\ref{nature1}) and (\ref{nature2}) in a self-consistent
manner as
\begin{eqnarray}
q_{\rm A}&=&q\left (1-\sum_{k} p_{\rm A}(k)I_{\rm A}^k \right ) \cr
&=& q\left (1-G_{\rm A}(1-q_{\rm B} + q_{\rm B} f_{\rm A} ) \right )
\label{nature3}
\end{eqnarray}
and
\begin{eqnarray}
q_{\rm B}= q\left (1-G_{\rm B}(1-q_{\rm A} + q_{\rm A} f_{\rm B} ) \right ),
\label{nature4}
\end{eqnarray}
respectively, where $G_{\rm A}(x)=\sum_{k} p_{\rm A}(k)x^k$ and
$G_{\rm B}(x)=\sum_{k}p_{\rm B}(k)x^k$. Equations
(\ref{nature1})--(\ref{nature4}) constitute a set of conditions for
determining four variables: $f_{\rm A}, f_{\rm B}, q_{\rm A}$, and
$q_{\rm B}$. Using these variables, the size of the mutual GC is
evaluated as
\begin{eqnarray}
&&\mu^{*}=q \left (1-G_{\rm A}(1-q_{\rm B} + q_{\rm B} f_{\rm A} ) \right )\cr
&& \hspace*{2cm} \times
\left (1-G_{\rm B}(1-q_{\rm A} + q_{\rm A} f_{\rm B} ) \right ).
\label{nature5}
\end{eqnarray}

For Erd\"{o}s-Renyi ensembles in particular,  which are
characterized by $p_{\rm A}(k)=e^{-a} a^k/k!$ and $p_{\rm
B}(k)=e^{-b} b^k/k!$, Eqs.~(\ref{nature1})--(\ref{nature4}) can be
summarized as two equations, since $G_{\rm A}(x)$ and $G_{\rm B}(x)$
accord to $H_{\rm A}(x)$ and $H_{\rm B}(x)$, respectively, as
$G_{\rm A}(x)=H_{\rm A}(x)=\exp(a(x-1))$ and $G_{\rm B}(x)=H_{\rm
B}(x)=\exp(b(x-1))$. The resultant coupled equations can be read as
\begin{eqnarray}
f_{\rm A}=\exp\left (-a(f_{\rm A}-1)(f_{\rm B}-1) \right ) \label{naturefA}
\end{eqnarray}
and
\begin{eqnarray}
f_{\rm B}=\exp\left (-b(f_{\rm A}-1)(f_{\rm B}-1) \right ),   \label{naturefB}
\end{eqnarray}
which are identical to Eq.~(14) in the Supplemental Material of
Ref. \cite{Buldyrev 2010}.

These two examples show that our scheme can be expressed compactly
using the generating functions when the macroscopic cavity variables
$I^{\rm A}_{l_{\rm A} l_{\rm B}}$ and $I^{\rm B}_{l_{\rm A} l_{\rm
B}}$ are independent of the degrees $l_{\rm A}$ and $l_{\rm B}$ as a
consequence of the assumed statistical features of objective systems
and failures. However, correlations of the degrees and/or site
dependence of the failures induces degree dependence of the
macroscopic cavity variables. In such cases, one has no choice but
to directly handle Eqs.~(\ref{macroh_stepA})--(\ref{mutualGC2t}) to
theoretically describe the behavior of interdependent networks.

\section{Numerical tests and results}
\label{section: numerical_test}
\subsection{The flow}
To confirm the validity of the analytical scheme, we carried out
numerical experiments using interdependent networks 
of $N=10000$
characterized by
a set of joint degree distributions $P(k_{\rm A},k_{\rm B})$,
$r_{\rm A}(k_{\rm A}, l_{\rm A})$, and $r_{\rm B}(k_{\rm B},l_{\rm
B})$ on the basis of the Monte Carlo algorithm proposed in
Ref. \cite{Newman 2002}. We measured the size of the GC, using the
algorithm proposed in Ref. \cite{Hoshen, Al-Futaisi}. To trigger the
cascade phenomena, sites in the constructed network A were removed
randomly (RF) or preferentially (TA).
In each case, we evaluated the
size of the mutual GC after convergence and compared it with the
analytical solution obtained by the cavity method.

As a simple but nontrivial example, we focused on 
two-peak models where 
the fractions of larger and smaller degrees are fifty-fifty.
We set the values of the larger and smaller degrees as $6$ and $4$, respectively, 
for both of networks A and B. 
In such models, the intranetwork degree--degree
correlations can be uniquely specified by the Pearson coefficient,
which is denoted as $C_{\rm A}$ and $C_{\rm B}$ for sub-networks A
and B, respectively. However, the influence of the internetwork
degree--degree correlations between networks A and B cannot be fully
characterized by only the Pearson coefficient of $P(k_{\rm A},k_{\rm
B})$, denoted as $C_{\rm I}$, because pairs of $(k_{\rm A}=4,k_{\rm
B}=6)$ and $(k_{\rm A}=6,k_{\rm B}=4)$ are affected by the initial
removal in different ways for TAs. However, for brevity, we suppose
that the fraction of these pairs are the same, which enables us to
characterize the intra- and internetwork degree--degree correlations
by three parameters" $(C_{\rm A},C_{\rm B},C_{\rm I})$.

The manner in which the failures occur also influences the size of
the mutual GC. In the case of RF, the randomness of the initial
removal means that the initial survival probability of the site
pairs does not depend on their degree pairs. For a TA, however, site
pairs that have a higher degree in A are removed preferentially, and
how the removal influences the sites in B in the first stage depends
on the internetwork degree--degree correlation. This implies that
the robustness of the interdependent networks can depend on the
internetwork degree--degree correlations in a nontrivial manner.


\subsection{Results and discussion}

\begin{figure}[htbp]
\begin{center}
\includegraphics[width=7cm,height=7cm]{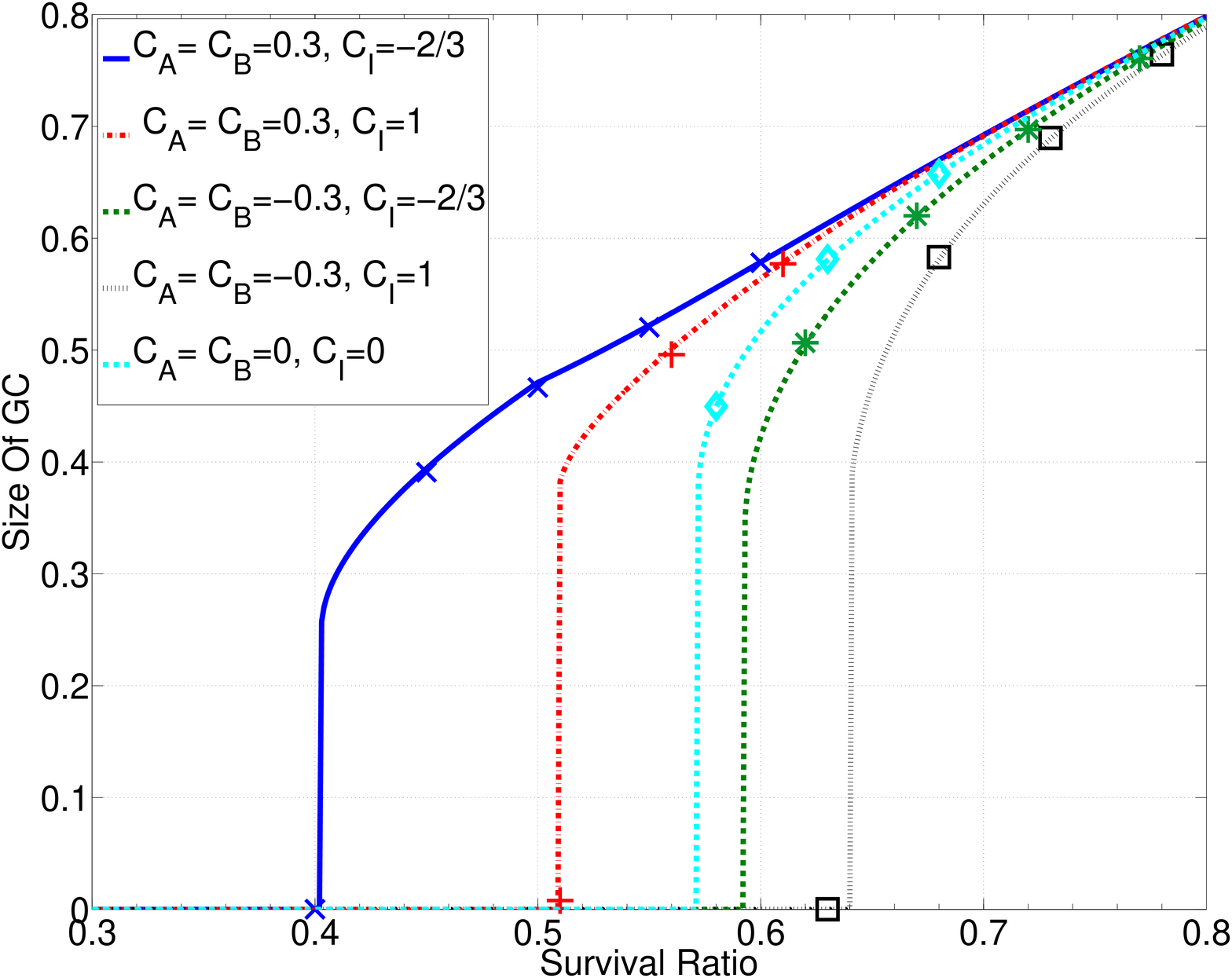}
\caption{
 \label{fig:TA}
(Color online) Cascade phenomena of two-peak interdependent
networks triggered by a TA. Each symbol represents the size of the
mutual GC evaluated from 50 experiments
for networks of $N=10000$.
Statistical errors are smaller than the size of the symbols. 
}
\end{center}
\end{figure}
\begin{figure}[htbp]
\begin{center}
\includegraphics[width=7cm,height=7cm]{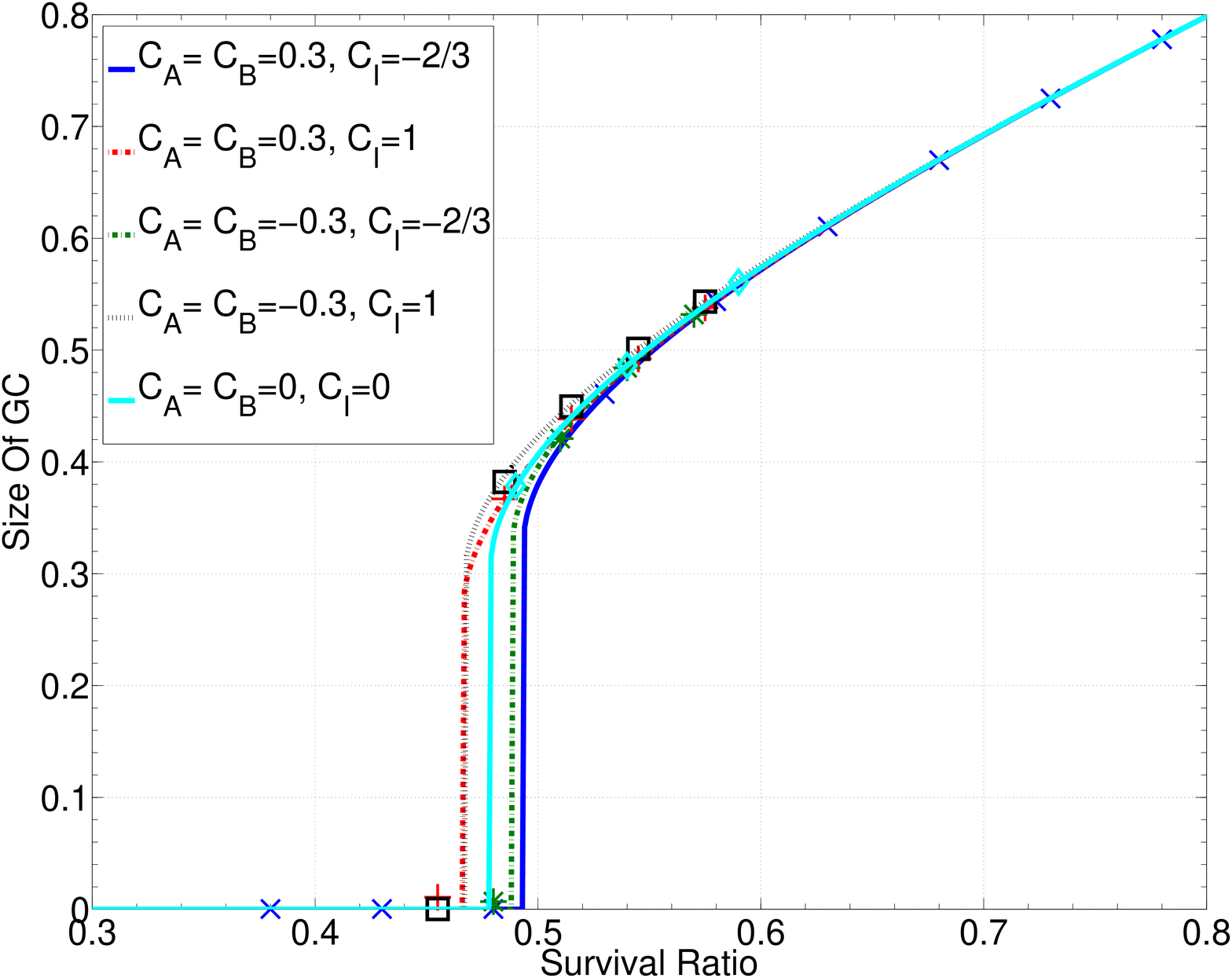}
\caption{\label{fig:RF}
(Color online) Cascade phenomena of two-peak interdependent networks
triggered by a RF. Each symbol represents the size of the mutual GC
evaluated from 50 experiments
for networks of $N=10000$.
Statistical errors are smaller than the size of the symbols. 
} 
\end{center}
\end{figure}
Figures \ref{fig:TA} and \ref{fig:RF} show how the size of the
mutual GC depends on the fraction of initial failures in the cases
of TAs and RFs, respectively. The solid lines represent estimates
obtained from the cavity method, while the symbols denote the
results of the numerical experiments. The results are in agreement
with an excellent accuracy, which validates our cavity-based
analytical scheme. The figures indicate that the percolation
transition of the interdependent networks remains discontinuous
irrespective of the introduction of the intra- and/or internetwork
degree--degree correlations.
\begin{figure}[htbp]
\begin{center}
\includegraphics[width=7cm,height=20cm]{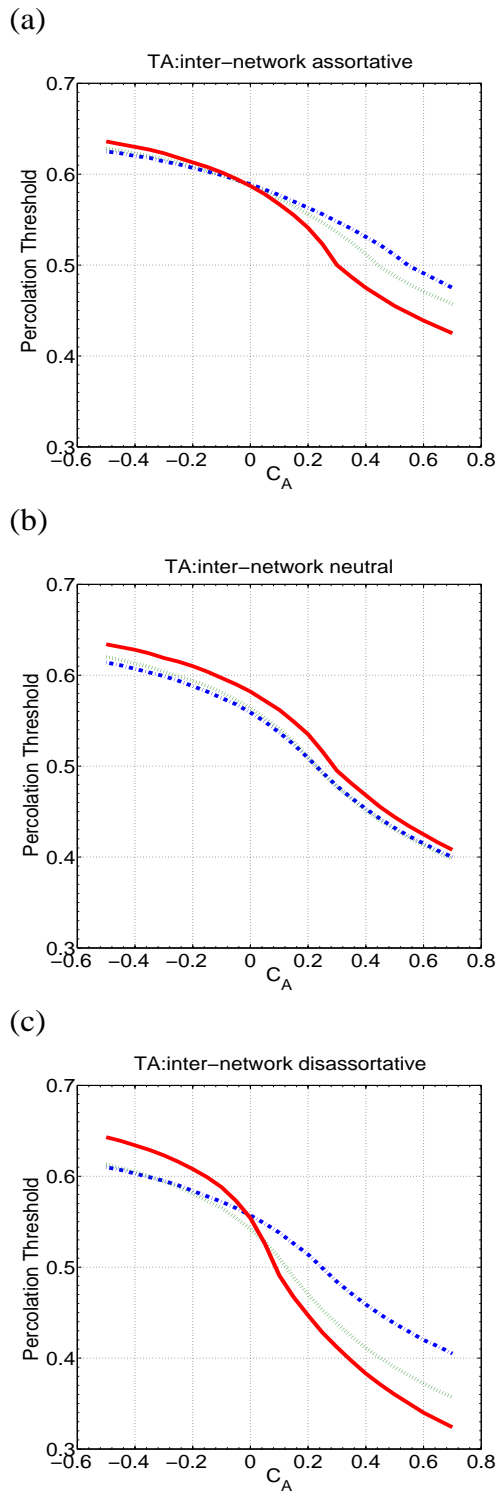}
\caption{(Color online) Percolation threshold as a function of the Pearson
coefficient of network A, $C_{\rm A}$, in the case of a TA; (a),
(b), and (c) correspond to three different values of the Pearson
coefficient for the internetwork degree--degree correlations $C_{\rm
I}=0.5$, $0$, and $-0.5$, respectively. Results are shown for
different values of the Pearson coefficient of network B $C_{\rm
B}$: $0.7$ (solid red line; assortative), $0$ (dashed green line;
neutral), and $-0.5$ (dash-dotted blue line: disassortative). }
\label{fig:shift_TA}
\end{center}
\end{figure}
\begin{figure}[htbp]
\begin{center}
\includegraphics[width=7cm,height=20cm]{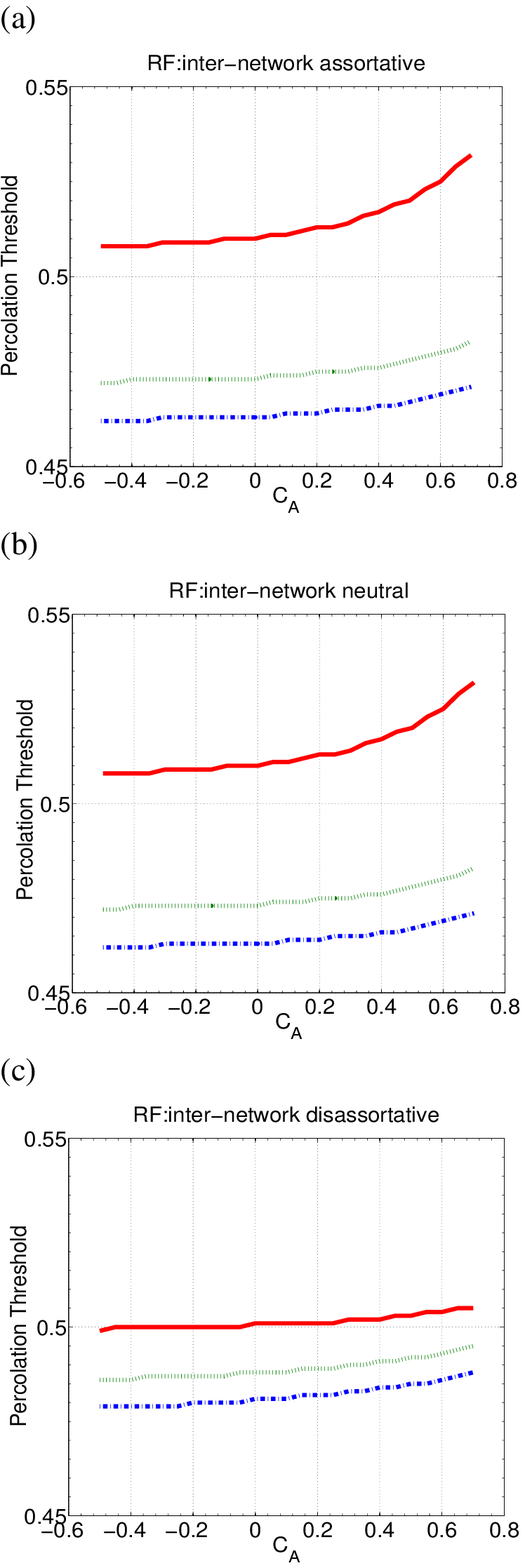}
\caption{(Color online) Percolation threshold as a function of the Pearson
coefficient of network A, $C_{\rm A}$, in the case of RF. The same
lines as those described in the legend of Fig.~\ref{fig:shift_TA} are used. }
\label{fig:shift_RF}
\end{center}
\end{figure}

Figures~\ref{fig:shift_TA} (a)--(c) show how the percolation
threshold depends on the intranetwork degree correlations for a TA.
The results indicate that the percolation threshold depends strongly
on the various degree correlations, and the interdependent networks
become more tolerant by introducing assortative intranetwork
degree--degree correlations to each network in the presence of the
internetwork degree--degree correlations, whether they are
assortative [Fig.~\ref{fig:shift_TA} (a)] or disassortative
[Fig.~\ref{fig:shift_TA} (c)]. In the absence of internetwork
degree--degree correlations [Fig.~\ref{fig:shift_TA} (b)], however,
the networks are the most tolerant when the intranetwork
degree--degree correlations in A are assortative and those in B are
disassortative.

A significantly different dependence on the intranetwork
degree--degree correlations is observed for RF.
Figures~\ref{fig:shift_RF} (a)--(c) indicate that whatever
type of internetwork degree--degree correlation is introduced,
disassortative intranetwork degree--degree correlations in both
networks A and B provide the highest robustness. However, as a
whole, the change in the percolation threshold as a function of the
various degree--degree correlations is relatively small. This
indicates that the role of the degree--degree correlations is not
significant in the case of RF, which is in contrast to the case of
TA.

Finally, we tested cases of various degree combinations 
using the cavity method. Tables \ref{table:RF} and \ref{table:TA} show the obtained 
values of the percolation threshold for TA and RF, respectively.
The results indicate the network robustness depends on 
the inter- and intranetwork correlations and the degree combinations in a complicated manner. 

First, we focus on the effects of the internetwork correlations. 
Table \ref{table:RF} indicates that the strong assortative internetwork correlations 
raise the robustness for RF lowering the threshold values except for the case of the 
minimal degree difference ($k_{\rm A,B}^1=5, k_{\rm A,B}^2=6$).  
Due to the interdependency, a site pair is disconnected from the mutual GC unless it belongs to the GCs in the both networks.
For RF, sites of the larger degrees are more likely to belong to GC in each sub-network. 
The  internetwork assortativity increases the fraction of the site pairs that have the larger degrees in both networks A and B. 
These imply that the probability that a site pair randomly picked up belongs to the mutual GC gets higher
as the internetwork assortativity is stronger and, therefore, 
the interdependent network becomes more robust.
However, the role of such effects becomes weaker as the degree difference is smaller.  
This is probably the reason why the strongest internetwork assortativity ($C_{\rm I}=1$) does not yield the highest robustness for 
the case of disassortative intranetwork correlations ($C_{\rm A}=C_{\rm B}=-0.3$) of the 
minimal degree difference.  

The robustness for RFs sometimes leads to the fragility for TAs in the case of single networks, e.g., scale free networks \cite{Barabasi}. 
Table \ref{table:TA} shows that this is also the case as a whole for the interdependent networks. 
However, the case of the minimal degree difference again exhibits an exceptional behavior.
This is supposed to be due to a similar reason as mentioned for RF. 


Next, let us turn to the influences of the intra-network correlations and  the degree combinations. 
Table \ref{table:RF} indicates that the network robustness for RF does not depend significantly 
on the intranetwork assortativity $(C_{\rm A}=C_{\rm B})$ for all degree combinations, 
which is consistent with the results suggested in Fig.~\ref{fig:shift_RF}. 
On the other hand, Table \ref{table:TA} shows that 
the robustness for TA can be influenced largely by the intranetwork assortativity. 
As a rule of thumb, the stronger intranetwork assortativity is likely to raise the robustness for TA, 
which however does not hold for the case of the large 
degree difference ($k_{\rm A,B}^1=3, k_{\rm A,B}^2=8$) and 
the strongest internetwork assortativity $(C_{\rm I}=1)$. 

The results obtained above imply that designing the most robust 
interdependent network taking into account the various degree--degree correlations is 
a highly nontrivial and challenging task.

\begin{table}
\begin{tabular}{|c|c|c|c|c|c|}\hline
$k_{\rm A}^1(k_{\rm B}^1)$ & $k_{\rm A}^2(k_{\rm B}^2)$ & $C_{\rm A}$ & $C_{\rm B}$ & $C_{\rm I}$ &survival ratio \\ \hline\hline
3& 8 & 0.3& 0.3& 1& 0.361\\ \hline
3& 8 & 0.3& 0.3& 0& 0.475\\ \hline
3& 8 & 0.3& 0.3& -1& 0.543\\ \hline
3& 8 & -0.3& -0.3& 1& 0.389\\ \hline
3& 8 & -0.3& -0.3& 0& 0.432\\ \hline
3& 8 & -0.3& -0.3& -1& 0.512\\ \hline
 \hline
4& 7 & 0.3& 0.3& 1& 0.413\\ \hline 
4& 7 & 0.3& 0.3& 0& 0.457\\ \hline 
4& 7 & 0.3& 0.3& -1& 0.466\\ \hline
4& 7 & -0.3& -0.3& 1& 0.419\\ \hline 
4& 7 & -0.3& -0.3& 0& 0.428\\ \hline 
4& 7 & -0.3& -0.3& -1& 0.458\\ \hline
 \hline
5& 6 & 0.3& 0.3& 1& 0.431\\ \hline 
5& 6 & 0.3& 0.3& 0& 0.444\\ \hline 
5& 6 & 0.3& 0.3& -1& 0.436\\ \hline
5& 6 & -0.3& -0.3& 1& 0.431\\ \hline 
5& 6 & -0.3& -0.3& 0& 0.426\\ \hline 
5& 6 & -0.3& -0.3& -1& 0.435\\ \hline
\end{tabular}
\caption{Percolation thresholds of two-peak interdependent networks for RF}
\label{table:RF}
\end{table}

\begin{table}
\begin{tabular}{|c|c|c|c|c|c|}\hline
$k_{\rm A}^1(k_{\rm B}^1)$ & $k_{\rm A}^2(k_{\rm B}^2)$ & $C_{\rm A}$ & $C_{\rm B}$ & $C_{\rm I}$ &survival ratio \\ \hline\hline
3& 8 & 0.3& 0.3& 1& 0.659\\ \hline
3& 8 & 0.3& 0.3& 0& 0.626\\ \hline
3& 8 & 0.3& 0.3& -1& 0.577\\ \hline
3& 8 & -0.3& -0.3& 1& 0.657\\ \hline
3& 8 & -0.3& -0.3& 0& 0.639\\ \hline
3& 8 & -0.3& -0.3& -1& 0.631\\ \hline
\hline
4& 7 & 0.3& 0.3& 1& 0.538\\ \hline 
4& 7 & 0.3& 0.3& 0& 0.495\\ \hline 
4& 7 & 0.3& 0.3& -1& 0.397\\ \hline
4& 7 & -0.3& -0.3& 1& 0.633\\ \hline 
4& 7 & -0.3& -0.3& 0& 0.6\\ \hline 
4& 7 & -0.3& -0.3& -1& 0.585\\ \hline
\hline
5& 6 & 0.3& 0.3& 1& 0.383\\ \hline 
5& 6 & 0.3& 0.3& 0&0.408 \\ \hline 
5& 6 & 0.3& 0.3& -1& 0.339\\ \hline 
5& 6 & -0.3& -0.3& 1& 0.582\\ \hline 
5& 6 & -0.3& -0.3& 0& 0.546\\ \hline
5& 6 & -0.3& -0.3& -1&0.550 \\ \hline
\end{tabular}
\caption{Percolation thresholds of two-peak interdependent networks for TA}
\label{table:TA}
\end{table}
\section{Summary}
\label{sec:summary} In summary, we have developed an analytical
methodology for evaluating the size of the mutual GC for basic
interdependent networks composed of two sub-networks A and B. The
methodology is based on the cavity method, which makes it possible
to evaluate the size of the GC against TAs and RFs by solving a set
of macroscopic nonlinear equations derived from a local tree
approximation in conjunction with the self-averaging property. We
have shown that the cavity-based methodology is reduced to the
widely known GFF in the absence of any degree correlations and that
solving the full cavity equations is indispensable for evaluating
the size of the GC in the presence of degree--degree correlations.

We compared the estimates of the size of the mutual GC with the
results of numerical experiments on two-peak degree distribution
models for site removal processes of TAs and RFs; there was
excellent consistency between the theory and experiments, which
validated the developed methodology. The utility of the methodology
was demonstrated by analyzing the degree correlation dependence of
the percolation threshold, which indicated that the network
robustness for TAs is sensitive to the intra- and internetwork
degree--degree correlations, whereas the significance of the
degree--degree correlations is relatively small for RFs.

Promising directions for future work include exploring the most
robust structure of an interdependent network system and more
general models that exemplify real-world systems.

\section*{Acknowledgements}
The authors thank  Koujin Takeda for useful comments and discussions. The authors also  appreciate anonymous refrees  whose remarks contribute greatly to the final version of the paper. This work was partially supported by JSPS/MEXT KAKENHI Grants No. 22300003, No. 22300098, and No. 25120013 (Y.K.).
Encouragement from the ELC project (Grant-in-Aid for
Scientific Research on Innovative Areas, JSPS/MEXT, Japan) is also acknowledged.
\bibliographystyle{tieice}
\bibliography{myrefs}

\end{document}